\documentclass[
    aps,
    fleqn,
    a4paper,
    superscriptaddress,
    preprint,
    preprintnumbers,
    floatfix,
    showkeys
]{revtex4}

\usepackage{graphicx}
\usepackage{epstopdf}
\usepackage{amsmath}
\usepackage{amsfonts}
\usepackage[utf8]{inputenc}
\usepackage{color}
\usepackage[
    colorlinks,
    linkcolor=blue,
    anchorcolor=blue,
    citecolor=blue,
    urlcolor=blue,
    menucolor=blue,
    filecolor=blue
]{hyperref}
\usepackage{subfigure}

\begin{document}


\title{Adaptive vibration suppression system: An iterative control law for a piezoelectric actuator shunted by a negative capacitor}

\author{Miloš  \surname{Kodejška}}
\affiliation{Institute of Mechatronics and Computer Engineering, Technical University of Liberec, CZ-46117 Liberec, Czech Republic}

\author{Pavel \surname{Mokrý}}
\email{pavel.mokry@tul.cz} 
\affiliation{Institute of Mechatronics and Computer Engineering, Technical University of Liberec, CZ-46117 Liberec, Czech Republic}
\affiliation{Research Centre for Special Optics and Optoelectronic Systems (TOPTEC), Sobotecká 1660, CZ-51101 Turnov, Czech Republic}

\author{Václav \surname{Linhart}}
\affiliation{Institute of Mechatronics and Computer Engineering, Technical University of Liberec, CZ-46117 Liberec, Czech Republic}

\author{Jan \surname{Václavík}}
\affiliation{Institute of Mechatronics and Computer Engineering, Technical University of Liberec, CZ-46117 Liberec, Czech Republic}
\affiliation{Research Centre for Special Optics and Optoelectronic Systems (TOPTEC), Sobotecká 1660, CZ-51101 Turnov, Czech Republic}

\author{Tomáš \surname{Sluka}}
\affiliation{Ceramics Laboratory, Swiss Federal Institute of Technology (EPFL), CH-1015 Lausanne, Switzerland}

\preprint{{\em IEEE Transactions on Ultrasonics, Ferroelectrics, and Frequency Control, vol. 59, no. 12, December 2012}}

\date{\today}

\begin{abstract}
An adaptive system for the suppression of vibration transmission using a single piezoelectric actuator shunted by a negative capacitance circuit is presented. It is known that using negative capacitance shunt, the spring constant of piezoelectric actuator can be controlled to extreme values of zero or infinity. Since the value of spring constant controls a force transmitted through an elastic element, it is possible to achieve a reduction of transmissibility of vibrations through a piezoelectric actuator by reducing its effective spring constant. The narrow frequency range and broad frequency range vibration isolation systems are analyzed, modeled, and experimentally investigated. The problem of high sensitivity of the vibration control system to varying operational conditions is resolved by applying an adaptive control to the circuit parameters of the negative capacitor. A control law that is based on the estimation of the value of effective spring constant of shunted piezoelectric actuator is presented. An adaptive system, which achieves a self-adjustment of the negative capacitor parameters is presented. It is shown that such an arrangement allows a design of a simple electronic system, which, however, offers a great vibration isolation efficiency in variable vibration conditions.
\end{abstract}


\maketitle

\section{Introduction}
\label{sec:intro}

Vibration suppression is nowadays regarded as an important issue in many technical fields ranging from robust aerospace industry to delicate nanotechnology. Unsuppressed vibrations are traditionally the key source of noise pollution, aging of mechanical components or even life-threatening fatal failures of transport vehicles, machining instruments, etc. Also in micro- and nanotechnology, the progress relies on high precision devices, which essentially require efficient vibration control. 

Contemporary vibration control techniques are based mostly on (i) passive methods using elements such as viscoelastic dampers and springs, or (ii) conventional active feedback and feed-forward control principles. The passive methods are rather inexpensive and do not require an external source of energy, but they are bulky and inefficient at low frequencies. On the other hand the conventional active methods can achieve an excellent efficiency with a subtle device, but on the expense of high technical complexity, high costs, and lower reliability.

There is necessarily a gap between the conventional passive and active vibration control methods, which would balance advantages of both approaches: especially a high efficiency (also at low frequencies) and a low cost. A promising new approach has emerged with so called semi-active methods, which have been heralded as the dawn of a new era in vibration control methods. However, it did not happen and yet most suggested semi-active methods remain confined in research laboratories.

In this article, we present a study of the semi-active vibration suppression method, which uses a piezoelectric bulk actuator. The vibration suppression effect is achieved: first, by inserting a piezoelectric actuator between vibrating structure and an object that is being isolated from vibrations, and, second, by connecting the piezoelectric actuator to an active external shunt circuit that controls the effective elastic stiffness of the actuator. The aforementioned method for a suppression of vibration transmission was introduced by Hagood and von Flotow \cite{Hagood1991} and, later \cite{Moheimani2006}, it was named as Piezoelectric Shunt Damping (PSD). During last two decades, extensive amount of work have been published on PSD method. Examples to be mentioned here are passive\cite{Tsai1999,Petit2004.proc} and active \cite{Morgan2002,Morgan2002.jva-tasme.124.77,RefLast1}, broadband multi-mode \cite{Petit2004.proc,Behrens2005.sms.12.18,Niederberger2004.sms.13.1025}, and, adaptive  \cite{Niederberger2004.sms.13.1025,Fleming2003.sms.12.36,Badel2006.jasa.119.2815} noise and vibration control devices. In a vast majority of the mentioned publications and many others, the classical control theory is used for a description and analysis of noise and vibration suppression systems. 

An alternative approach was offered by Date et al. \cite{Date2000}, who discovered that the effect of the shunt circuit on the mechanical response of the piezoelectric actuator can be explained through the change of  effective elastic properties of a piezoelectric actuator. When an actuator is inserted between a source of vibrations and an object that is being isolated from vibrations, the resonant frequency and the transmissibility of vibrations through a resulting spring-mass system depends on the spring constant of the actuator and the mass of the object. The reduction of the piezoelectric actuator spring constant results in the reduction of resonant frequency and transmissibility of vibrations at super-resonant frequencies. Therefore, the physics lying behind the vibration suppression effect in PSD method is in principle the same as in the passive methods. On top of that, PSD method enables the efficient vibration suppression at low frequencies, too. 

Such a principal change in the concept offers a use of alternative and simpler design tools in the development of noise and vibration suppression devices. The design of vibration control systems can be reduced to, first, the study of elasticity effect on the noise or vibration transmission and, second, to the realization of Active Elasticity Control. Early applications that followed this simple approach  \cite{Mokry2003jul,Mokry2003dec,Imoto2005,Tahara2006,Kodama2008} demonstrated the great potential of this method, which stems from (i) the simplicity of the noise control system, which consists of a self-sensing piezoelectric actuator connected to an active shunt circuit, (ii) an implementation of active shunt circuit electronics using a simple analog circuit with a single linear power amplifier that allows a significant reduction of the electric power consumption, and (iii) the broad frequency range (e.g. from 10 Hz to 100 kHz) where the system can efficiently suppress vibrations.

Despite the potential advantages of the PSD method, high sensitivity and low stability of real systems currently prevents their industrial exploitation, which often requires a system robustness under varying operational conditions. It was actually shown by Sluka et al. \cite{Sluka2007} that the optimal working point of the PDS system lies just on the edge of the system stability. Later, the stability of several PSD method implementations has been compared in the work by Preumont et al. \cite{Preumont2008}. A partial elimination of the drawbacks was achieved by adaptive PSD vibration control systems reported in Refs. \cite{Sluka2007,Sluka2008}, however, with a strong limitation that the stable vibration isolation efficiency could be achieved only in a narrow frequency range.

The aforementioned issues have motivated the work presented below, where we will address the design of an adaptive broad-band vibration control device. The principle of our vibration suppression device will be presented in Sec. \ref{sec:principle}. In Sec. \ref{sec:manual} we will demonstrate advantages and drawbacks of a narrow and broad frequency range vibration isolation device with manually adjusted negative capacitor. Design of the adaptive broad-band vibration isolation device will be presented in Sec. \ref{sec:adaptive}. There, in addition, we will present previously unpublished details of control law, which was used to obtain results presented in Refs.~\cite{Sluka2007,Sluka2008}. Conclusions of our experiments will be presented in Sec.~\ref{sec:conclusions}.

\section{Principle of the vibration suppression}
\label{sec:principle}

It is known that the vibration transmission through an interface between two solid objects is mainly controlled by the ratio of their mechanical impedances. Since the mechanical impedance is proportional to the material stiffness, extremely soft element placed between two other objects works as an interface with high transmission loss of vibrations. In the following Subsection, we present a simple theoretical model that explains the effect of the elasticity in a mechanical system on the transmission of vibrations through the system. Later, we present a method to control the elastic properties of the piezoelectric actuator using a shunt electric circuit that can be profitably used in the vibration isolation system.

\begin{figure}[t]
\begin{center}
\includegraphics[width=85mm]{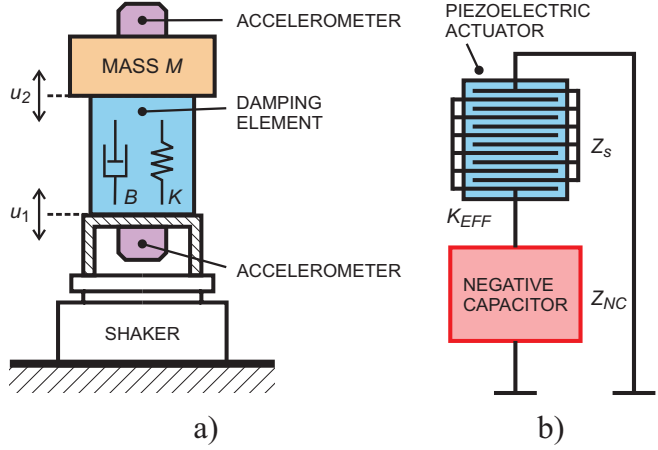}
\caption{
Scheme of the vibration isolation measurement system. The vibration damping element with a spring constant $K$ and a damping coefficient $B$ are placed between the shaker and a mass $M$ that is going to be isolated from vibrations. The incident vibrations of a displacement amplitude $u_1$ and the transmitted vibrations of a displacement amplitude $u_2$ are measured using accelerometers (a). The vibration damping element used in this work is the piezoelectric actuator of the impedance $Z_S$ shunted by a negative capacitor of the impedance $Z_{\rm NC}$. (b) 
}
    \label{fig:01-scheme}
\end{center}
\end{figure}

\subsection{Effect of the spring constant on the transmissibility of vibrations}
\label{subsec:tr_spring_effect}

Scheme of the vibration isolation measurement system is shown in Fig.~\ref{fig:01-scheme}(a). The vibration damping element with a spring constant $K$ and a damping coefficient $B$ is placed between the shaker and the object of a mass $M$ that is going to be isolated from vibrations. The incident and transmitted vibrations of a displacement amplitudes $u_1$ and $u_2$, respectively, are measured using accelerometers. The transmissibility of vibrations $TR$ through the considered vibration isolation system is defined as a ratio of the transmitted over the incident displacement amplitudes at the reference source point:
\begin{equation}
	\mathit{TR}=|u_2/u_1|
	\label{eq:01}
\end{equation}

The transmissibility of vibration is a function of material parameters that control the dynamic response of the mechanical system. The dynamic response of the system is governed by the following equation of motion:
\begin{equation}
M\frac{d^{2}u_{2}}{\mathit{dt}^{2}}+B\frac{du_{2}}{\mathit{dt}}+Ku_{2}=B\frac{du_{1}}{\mathit{dt}}+Ku_{1}.
\label{eq:02}
\end{equation}
Considering the simplest case of the transmission of harmonic vibrations of an angular frequency ${\omega}$, the solution of Eq. (\ref{eq:02}) yields the formula:
\begin{equation}
	\mathit{TR}=\mathit{\omega\,}_{0}\sqrt{\frac{\mathit{\omega\,}^{2}+Q^{2}\omega\,_{0}^{2}}{\omega\,^{2}\omega\,_{0}^{2}+Q^{2}\left(\omega\,_{0}^{2}-\omega\,^{2}\right)^{2}}},
	\label{eq:03}
\end{equation}
where the symbols $Q$ and  $\omega\,_{0}$  \ stand for the mechanical quality factor  $Q=\sqrt{\mathit{KM}}/B$ and the resonance frequency  $\omega\,_{0}=\sqrt{K/M}$. It is seen that the smaller the value of spring constant $K$, the smaller the value of the resonant frequency  $\omega\,_{0}$, and the smaller the value of transmissibility $TR$ of harmonic vibrations of angular frequency  $\omega\,>\omega\,_{0}$.

\subsection{Method of the active control of piezoelectric actuator elasticity}
\label{subsec:AEC}

Figure~\ref{fig:01-scheme}(b) shows the vibration damping element used in this work, which is a piezoelectric actuator of capacitance $C_{S}$ shunted by a negative capacitor of capacitance $C$. This system is an example of so called Active Elasticity Control method introduced in 2000 by Date et al. \cite{Date2000}. The effective spring constant of a piezoelectric actuator $K_{\rm eff}$ can be derived from the constitutive equations for charge $Q$ and change of length $\Delta l=u_{2}-u_{1}$ of a piezoelectric actuator:
\begin{eqnarray}
	\label{eq:04}
	Q&=&\mathit{dF}+C_{S}V, \\
	\label{eq:05}
	\Delta l&=&(1/K_{S})F+dV,
\end{eqnarray}
which are appended by the formula for a voltage $V$ applied back to the piezoelectric actuator from a shunt circuit of capacitance $C$:
\begin{equation}
V=-Q/C,
	\label{eq:06}
\end{equation}
where symbols $d$, $C_{S}$, and $K_{S}$ stand for the piezoelectric coefficient, capacitance, and spring constant of a mechanically free piezoelectric actuator, respectively.

Combining Eqs. (\ref{eq:04}), (\ref{eq:05}), and (\ref{eq:06}) and with the use of relationship between the capacitance and impedance of a capacitor,  $Z=1/(j\omega\,C),$ one can readily obtain the formula for the effective spring constant of a piezoelectric actuator connected to an external shunt circuit with an electric impedance $Z$:
\begin{equation}
	K_{\rm eff}
		=\frac{F}{\Delta l}
		=K_{S}\left(\frac{1+Z_{S}/Z}{1-k^{2}+Z_{S}/Z}\right),
	\label{eq:07}
\end{equation}
where $k^{2}=d^{2}K_{S}/C_{S}$ is the electromechanical coupling factor of the piezoelectric element $(0<k<1)$ and $Z_S$ is the electric impedance of a mechanically free piezoelectric actuator. 

\begin{figure}[t]
\begin{center}
    \includegraphics[width=60mm]{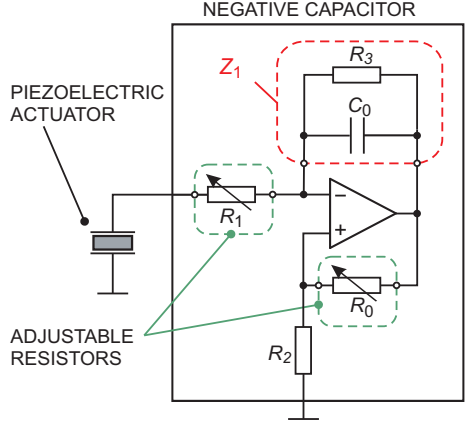}
\caption{
Electrical scheme of the piezoelectric actuator shunted by a negative capacitor. The negative capacitor is designed using a simple circuit with an operational amplifier in a feedback loop. By use of adjustable resistors  $R_{0}$  and  $R_{1}$, it is possible to adjust the real and imaginary part of its capacitance, so that it matches the capacitance of the piezoelectric actuator (except the sign).
}
    \label{fig:02-NCscheme}
\end{center}
\end{figure}
%
It follows from Eq. (\ref{eq:07}) that, when the complex impedance of the shunt circuit $Z$ approaches the value of $-Z_{S}$, the effective spring constant $K_{\rm eff}$ of the piezoelectric element reaches zero. Figure~\ref{fig:02-NCscheme} shows the electrical scheme of the piezoelectric actuator shunted by the active circuit that effectively works as with a negative capacitance. It will be further referenced as a negative capacitor. Effective impedance of the negative capacitor shown in Fig.~\ref{fig:02-NCscheme} is equal to
\begin{equation}
Z(\omega\,)=R_{1}+\frac{R_{0}+R_{2}+A_{u}(\omega\,)R_{2}}{R_{0}+R_{2}-A_{u}(\omega\,)R_{0}}Z_{1}(\omega\,)\approx R_{1}-\frac{R_{2}}{R_{0}}Z_{1}(\omega\,),
	\label{eq:08}
\end{equation}
where $A_u$ is the output voltage gain of the operational amplifier and
\begin{equation}
Z_{1}(\omega\,)=\frac{R_{3}}{1+j\omega\,C_{0}R_{3}}=\frac{R_{3}-j\omega\,C_{0}R_{3}^{2}}{1+\omega\,^{2}C_{0}^{2}R_{3}^{2}}
	\label{eq:09}
\end{equation}
is the impedance of so called reference capacitance of the negative capacitor. The approximate formula on the right-hand side of Eq.~(\ref{eq:08}) stands for the ideal operational amplifier, i.e. $A_u$ goes to infinity. 

It is known that real and imaginary parts of the piezoelectric actuator capacitance practically do not depend on frequency in the frequency range below its resonance frequency. In this situation, the capacitance of the piezoelectric actuator can be approximated with a high accuracy by the formula $C'_{S}(1-j\tan\delta_{S})$, where $C'_{S}$ and $\tan{\delta}_S$ are the real part and loss tangent of the piezoelectric actuator capacitance. Then, the impedance of the piezoelectric actuator is equal to 
\begin{equation}
Z_{S}(\omega\,)=\frac{1}{j\omega\,C'_{S}(1-j\tan \delta_{S})}=\frac{\tan \delta_{S}-j}{\omega\,C_{S}(1+\tan ^{2}\delta_{S})}.
	\label{eq:10}
\end{equation}
It is convenient to approximate the frequency dependence of the piezoelectric actuator impedance by the frequency dependence of the in-series connection of the capacitor and resistor of capacitance  $C_{S}$  and resistance $R_S$, respectively. 
\begin{equation}
Z_{S}(\omega\,)\approx R_{S}+\frac{1}{j\omega\,C_{S}}.
	\label{eq:11}
\end{equation}

At given critical frequency  $\omega\,_{0}$, it is possible to adjust the negative capacitor in such a way that:
\begin{subequations}
\label{eq:12}
\begin{eqnarray}
	\label{eq:12a}
\left|Z\right|(\omega\,_{0})&=&\left|Z_{S}\right|(\omega\,_{0}),\\
	\label{eq:12b}
\mathit{arg}[Z(\omega\,_{0})]&=&\mathit{arg}[Z_{S}(\omega\,_{0})]+\pi.
\end{eqnarray}
\end{subequations}
Such a situation is characterized by the relation $Z_{S}(\omega\,_{0})/Z(\omega\,_{0})=-1$ and, according to Eq.~(\ref{eq:07}), it yields $K_{\rm eff}$ that is effectively reaching zero value and the transmission of vibrations reaches minimum. 

\subsection{Role of impedance matching}
\label{subsec:role_of_z_matching}

In this subsection, we analyze to what extent the condition given by Eqs.~(\ref{eq:12}) must be satisfied, in order to achieve required suppression of vibration transmission. Such an analysis can be split into two steps. First, we analyze the sensitivity of transmissibility $TR$ to the value of spring constant of the actuator $K$, and, second, we analyze the sensitivity of the effective spring constant of the actuator $K_{\rm eff}$ to the capacitance $C$ of the negative capacitor.

\begin{figure*}[t]
\begin{center}
    \includegraphics[width=0.8\textwidth]{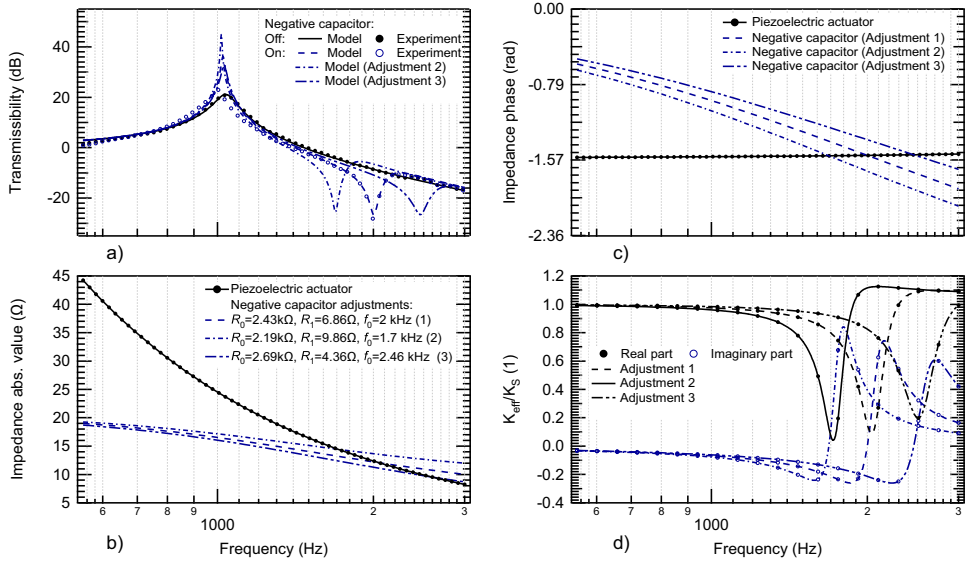}
\caption{
Frequency dependences of the physical quantities that controls the value of transmissibility of vibrations through the piezoelectric actuator shunted by the negative capacitor shown in Fig.~\ref{fig:02-NCscheme}: a) shows the comparison of the measured values of the transmissibility of vibrations through the electrically free piezoelectric actuator (filled circles) and the piezoelectric actuator shunted by the negative capacitor adjusted at the frequency  $f_{0}$  = 2 kHz (empty circles); the measured values of the transmissibility of vibrations are compared with the calculated ones from the theoretical model; b) absolute value of the electric impedance of the piezoelectric actuator (measured) and the negative capacitor for three different adjustments of resistors  $R_{0}$  and  $R_{1}$  (calculated); c) phase of the electric impedance of the piezoelectric actuator (measured) and of the negative capacitor (calculated and subtracted by $\pi$); d) calculated real and imaginary parts of the effective spring constant of the piezoelectric actuator shunted by the negative capacitor.}
\label{fig:03-ExpNarrow}
\end{center}
\end{figure*}

In order to perform the first step of the analysis, it is convenient to express the suppression level of the transmissibility of vibrations $\Delta L_{\rm TR}$, which is produced by the active elasticity control using the negative capacitor:
\begin{equation}
	\Delta L_{\rm TR} = 20\left(\log TR_{NC}- \log TR_{S}\right),
	\label{eq:13}
\end{equation}
where $TR_{NC}$ and $TR_{S}$ are the transmissibility of vibrations given by Eq.~(\ref{eq:03}) in situations, where the shunt circuit is connected and disconnected from the piezoelectric actuator, respectively. For small values of spring constant $K$ and for frequencies above the resonant frequency of the system $\omega_0$, it is possible to express the suppression level of the transmissibility of vibrations in the form:
\begin{equation}
	\Delta L_{\rm TR} \approx 10\,\log\left|K_{\rm eff}/K_S\right|,
	\label{eq:14}
\end{equation}
where the $K_{\rm eff}$ is the effective spring constant of the actuator, which is controlled by the negative capacitor. 

In the second step, it is convenient to denote $\Delta Z=Z-(-Z_S)$ as a deviation of the impedance $Z$ of the negative capacitor from the required value $-Z_S$. Then for small deviations $\Delta Z$, it is possible to approximate Eq. (\ref{eq:07}) by a formula:
\begin{equation}
	K_{\rm eff}\approx K_S \Delta Z/\left(k^2 Z_s\right)
	\label{eq:15}
\end{equation}
From Eq.~(\ref{eq:14}), it can be estimated that a decrease in the level of transmissibility of vibrations $\Delta L_{\rm TR}$ by about 20\,dB requires a decrease in the effective value of the spring constant $K$ by a factor of 1/100. Then, considering the values of the electromechanical coupling factor of conventional piezoelectric ceramics, i.e. $k^2 = 0.1$, one can conclude from Eq.~(\ref{eq:15}) that the relative deviation of the negative capacitor impedance $\delta Z=\Delta Z/Z_S$ from its required value $-Z_S$ must be smaller than 0.1\%. This very narrow region of capacitances of the negative capacitor, in which the required values of the spring constant are achieved, imposes high requirements on the negative capacitor adjustment. 

It is clear that required adjustment of the negative capacitor cannot be achieved with fixed values of resistors $R_0$ and $R_1$ due to  limited number of commercially available values and the continuously adjustable trimmers must be used. 
%
%
In the next two Sections, there are presented experimental results acquired on the vibration isolation systems with manual and adaptive adjustment of the negative capacitor

\section{Manual adjustment of the negative capacitor}
\label{sec:manual}

In the next subsection, we will present and discuss the experimental data measured on the vibration isolation device with the negative capacitor shown in Fig.~\ref{fig:02-NCscheme}. 

\subsection{Narrow frequency range vibration isolation}
\label{subsec:manual_narrow}

First, the frequency dependence of the transmissibility of vibrations through the electrically free piezoelectric actuator, i.e. the actuator, which was disconnected from the negative capacitor, was measured in the frequency range from 550~Hz to 3~kHz and the result is indicated by filled circles in Fig.~\ref{fig:03-ExpNarrow}(a). The measured frequency dependences of the transmissibility of vibrations were compared with predictions of the theoretical formula given by Eq. (\ref{eq:03}). Values of spring constant $K_{S}$ ~=~$7.11\cdot{}10^7$~Nm${}^{-1}$, mass $M$~=~1.67~kg and the mechanical quality factor of the piezoelectric actuator $Q$~=~11.3 were obtained using the method of least squares. 

In the next step, the negative capacitor was assembled using LF~356N operational amplifier according to scheme shown in Fig.~\ref{fig:02-NCscheme}. The output voltage gain of LF~356N was approximated by the function $A_{u}(\omega_{0})=A_{0}/(1+j\omega/(2\pi f_{1}))$, where $A_{0}$~=~105~dB and $f_{1}$~=~100~Hz. The condition given by Eqs.~(\ref{eq:12}) is achieved by setting the values of resistances $R_{1}$ and $R_{0}$ according to following formulae:
\begin{subequations}
\label{eq:16}
\begin{eqnarray}
	\label{eq:16a}
	R_{0}&=&\frac{\omega\,_{0}^{2}C_{0}C_{S}R_{2}R_{3}^{2}}{1+\omega\,^{2}C_{0}^{2}R_{3}^{2}}, \\
	\label{eq:16b}
	R_{1}&=&\frac{1}{\omega\,^{2}C_{0}C_{S}R_{3}}-R_{S}.
\end{eqnarray}
\end{subequations}

In order to find the proper adjustment of the negative capacitor, the frequency dependence of the electric impedances of the piezoelectric actuator and the reference capacitance $Z_{1}$ were measured using HP 4195A spectrum analyzer and shown in Figs.~\ref{fig:03-ExpNarrow}(b) and (c). Using the least squares method, following values were obtained: $R_{S}$~=~1.150~${\Omega}$, $C_{S}$~=~6.602~${\mu}$F, $R_{3}$~=~27.84~${\Omega}$, and $C_{0}$~=~4.686~${\mu}$F. These values were cross-checked by direct measurements on ESCORT ELS-3133A LRC-meter at 1~kHz: $R_{S}$~=~0.87~${\Omega}$, $C_{S}$~=~6.94~${\mu}$F, $R_{3}$~=~24.5~${\Omega}$, and $C_{0}$~=~5.16~${\mu}$F. Then, resistance $R_{2}$~=~2.40~k${\Omega}$ was measured and the negative capacitor resistors were pre-adjusted to values $R_{0}$~=~2.41~k${\Omega}$ and $R_{1}$~=~6.93~${\Omega}$ according to Eqs.~(\ref{eq:16}).

\begin{figure}[t]
\begin{center}
    \includegraphics[width=60mm]{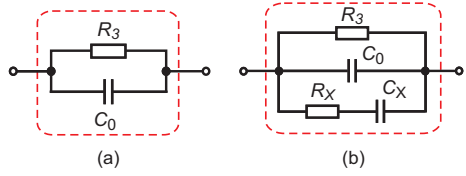}
\caption{
Electrical scheme of the reference impedance $Z_{1}$ inside the negative capacitor, which is shown in Fig.~\ref{fig:02-NCscheme}, for a narrow frequency range (a) and a broad frequency range vibration isolation system (b). }
\label{fig:04-Z1AandB}
\end{center}
\end{figure}
%

Afterwards, the trimmers $R_{0}$ and $R_{1}$ in the negative capacitor were finely tuned in order to achieve 20~dB decrease in the transmissibility of vibration at 2~kHz, as indicated by empty circles in Fig.~\ref{fig:03-ExpNarrow}(a). The measured transmissibility of vibrations was fitted to the theoretical model given by Eqs. (\ref{eq:03}), (\ref{eq:08}), (\ref{eq:09}), and (\ref{eq:11}). Following values were obtained using the method of least squares: $k^{2}$~=~0.064, $R_{0}$~=~2.43~k${\Omega}$, and $R_{1}$~=~6.86~${\Omega}$. The direct measurement, using the LRC-meter, resulted in following values: $R_{0}$~=~2.32~k${\Omega}$ and $R_{1}$~=~6.20~${\Omega}$, respectively. 

Here, it should be noted that the relative difference between fitted and measured values of resistances varies from 5\% to 11\%. This relative difference is much larger than 0.1\% allowed relative difference between the negative capacitor and piezoelectric actuator capacitances. The reason for such a difference is the presence of parasitic capacitances in the system, which makes theoretical modeling of piezoelectric shunt damping systems difficult and adjustment of negative capacitors from such theoretical models practically impossible. 

\begin{figure*}[t]
\begin{center}
    \includegraphics[width=0.8\textwidth]{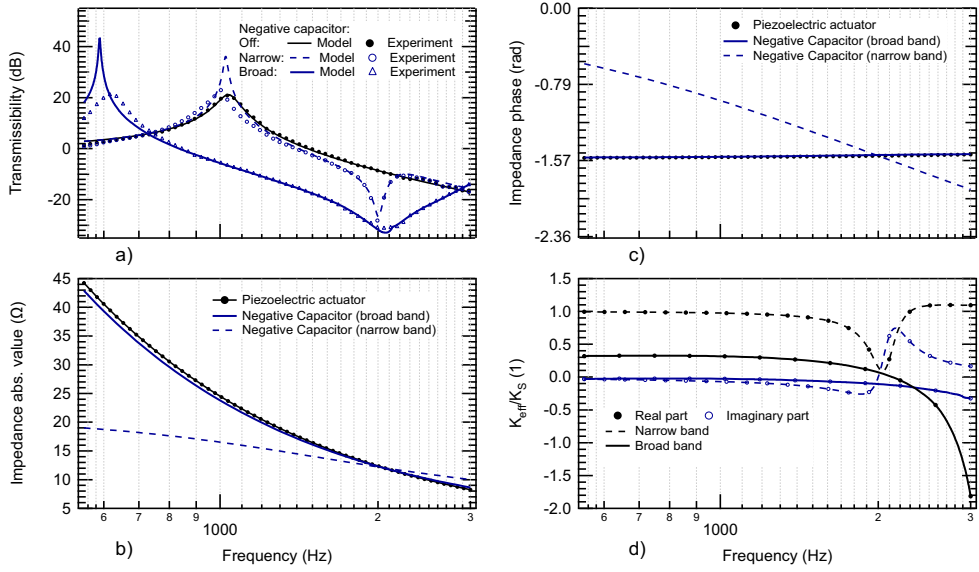}
\caption{
Frequency dependences of the physical quantities that control the value of transmissibility of vibrations through the piezoelectric actuator shunted by the negative capacitor shown in Fig.~\ref{fig:02-NCscheme}: a) shows the comparison of the measured values of the transmissibility of vibrations through the electrically free piezoelectric actuator (filled circles), through the piezoelectric actuator shunted by the narrow frequency range negative capacitor adjusted at $f_{0}$~=~2~kHz (empty circles), and the broad frequency range negative capacitor adjusted at $f_{0}$~=~2~kHz (empty triangles). The measured values of the transmissibility of vibrations are compared with the theoretical model; b) absolute value of the electric impedance of the piezoelectric actuator (measured) and the negative capacitor for the narrow frequency range (see Fig.~\ref{fig:04-Z1AandB}a) and broad frequency range (see Fig.~\ref{fig:04-Z1AandB}b) reference impedance  $Z_{1}$; c) phase of the electric impedance of the piezoelectric actuator (measured) and the negative capacitor (calculated and subtracted by $\pi$); d) calculated real and imaginary parts of the effective spring constant of the piezoelectric actuator shunted by the negative capacitor.}
\label{fig:05-broad}
\end{center}
\end{figure*}
%

The physics standing behind the decrease in the transmissibility of vibrations in a narrow frequency range can be easily understood by looking at Fig.~\ref{fig:03-ExpNarrow}. Figures~\ref{fig:03-ExpNarrow}(b) and (c) show the comparison of the measured electric impedance absolute value and phase of the piezoelectric actuator with the calculated values of the electric impedance absolute value and phase of the negative capacitor for three adjustments that differ in values of resistances $R_0$, $R_1$, and the critical frequency $f_0$. Figures~\ref{fig:03-ExpNarrow}(b) and (c) indicate that conditions given by Eqs.~(\ref{eq:12}) are satisfied only in narrow frequency ranges around particular critical frequencies $f_0$. This is a reason for narrow frequency ranges, where a decrease in the real part of the effective spring constant $K_{\rm eff}$ of the piezoelectric actuator can be achieved as indicated in Fig.~\ref{fig:03-ExpNarrow}(d).
%

The next Subsection discusses the problem of broadening the frequency range where the vibration isolation device can efficiently suppress the vibration transmission. 

\subsection{Broad frequency range vibration isolation}
\label{subsec:manual_broad}

In order to broaden the frequency range of the efficiently suppressed vibration transmission, it is necessary to achieve a precise matching the electrical impedances of the piezoelectric actuator and the negative capacitor. Since the frequency dependence of the piezoelectric actuator is controlled by its material and its construction, it is necessary to modify the frequency dependence of the negative capacitor. The frequency dependence of the negative capacitor impedance is determined by the reference impedance $Z_{1}$. The trivial parallel connection of the capacitor $C_{0}$ and the resistor $R_{3}$, which is shown in Fig.~\ref{fig:04-Z1AandB}(a), was replaced by a more complicated RC network shown in Fig.~\ref{fig:04-Z1AandB}(b).

\begin{figure}[t]
\begin{center}
    \includegraphics[width=82mm]{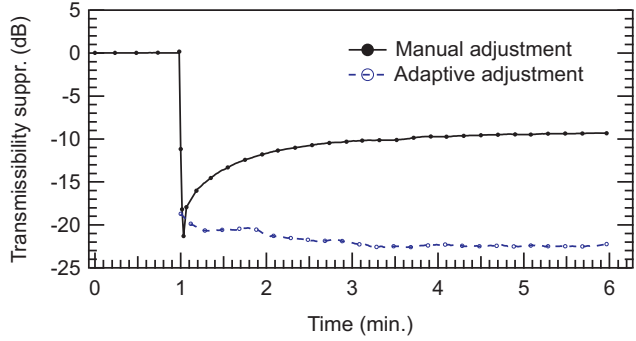}
\caption{
Comparison of the time dependences of the vibration isolation efficiency in changing operational conditions in the system with manually adjusted negative capacitor (solid line) and in the system with adaptively controlled negative capacitor (dashed line), respectively. The vibration isolation system was turned on at the time 1 min. Then, the ambient temperature of the system was changed. There is seen approximately 15 dB decrease in the suppression level of the transmissibility of vibrations [see Eq. (\ref{eq:13})] after 5 minutes in the system with manually adjusted negative capacitor, while the suppression level of the transmissibility of vibrations remains constant in the adaptive vibration isolation system.}
\label{fig:06-TrTimeManVsAuto}
\end{center}
\end{figure}
%

Values of capacitances and resistances in the reference impedance $Z_{1}$ were adjusted to minimize the mismatch between values of $Z_1$ and $Z_S$ in the frequency range from 0.5~kHz to 3~kHz. The frequency dependence of the electric impedance of the modified reference capacitance $Z_{1}$ was measured and the method of least squares yields the values: $R_{3}$~=~15.09~k${\Omega}$, $C_{0}$~=~480~nF, $R_{X}$~=~44.6~${\Omega}$, and $C_{X}$~=~807~nF. The fitted values were cross-checked by a direct measurements using LRC-meter at 1~kHz giving values: $R_{3}$~=~15~k${\Omega}$, $C_{0}$~=~470~nF, $R_{X}$~=~44~${\Omega}$, and $C_{X}$~=~813~nF.

Then, the trimmers $R_{0}$ and $R_{1}$ in the negative capacitor were finely tuned in order to achieve the maximum decrease in the transmissibility of vibrations at the frequency 2~kHz. Then the transmissibility of vibration through the piezoelectric actuator shunted by a broad-frequency-range-optimized negative capacitor was measured and the result is indicated by empty triangles in Fig.~\ref{fig:05-broad}(a). It can be seen that a 20~dB decrease in the transmissibility of vibration was achieved in the broad frequency range from 1~kHz to 2~kHz. The measured values of the frequency dependence of the transmissibility of vibrations were compared with the theoretical model given by Eqs. (\ref{eq:03}), (\ref{eq:08}), (\ref{eq:09}) and (\ref{eq:11}), and the values of $k^{2}$~=~0.067, $R_{0}$~=~12.6~k${\Omega}$, and $R_{1}$~=~2.6~${\Omega}$ using the method of least squares.

\begin{figure*}[t]
\centering
	\subfigure[$\left|K^\prime_{\rm eff}+iK^{\prime\prime}_{\rm eff}\right|$]{
	\includegraphics[width=76mm]{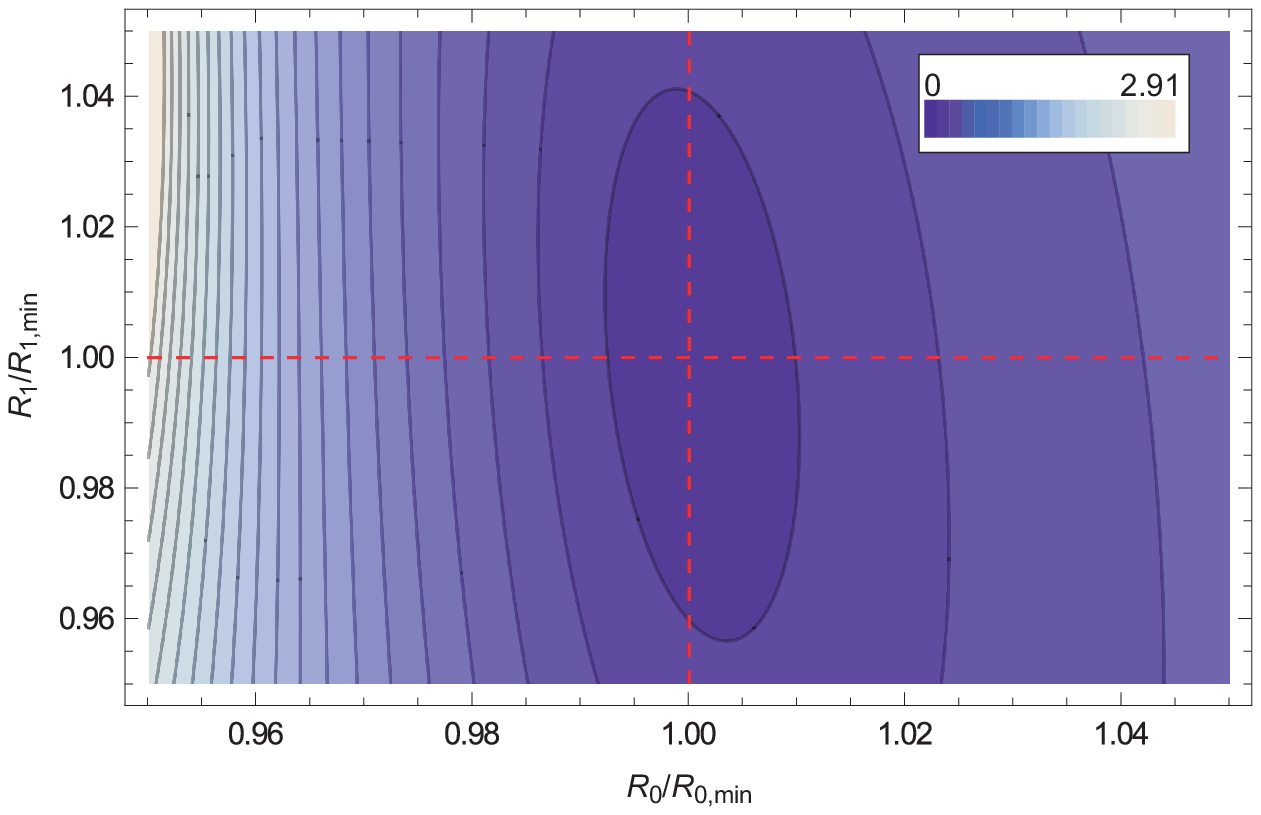}
	\label{fig:07-a}
	}
	\subfigure[$\arg\left(K^\prime_{\rm eff}+iK^{\prime\prime}_{\rm eff}\right)$]{
	\includegraphics[width=76mm]{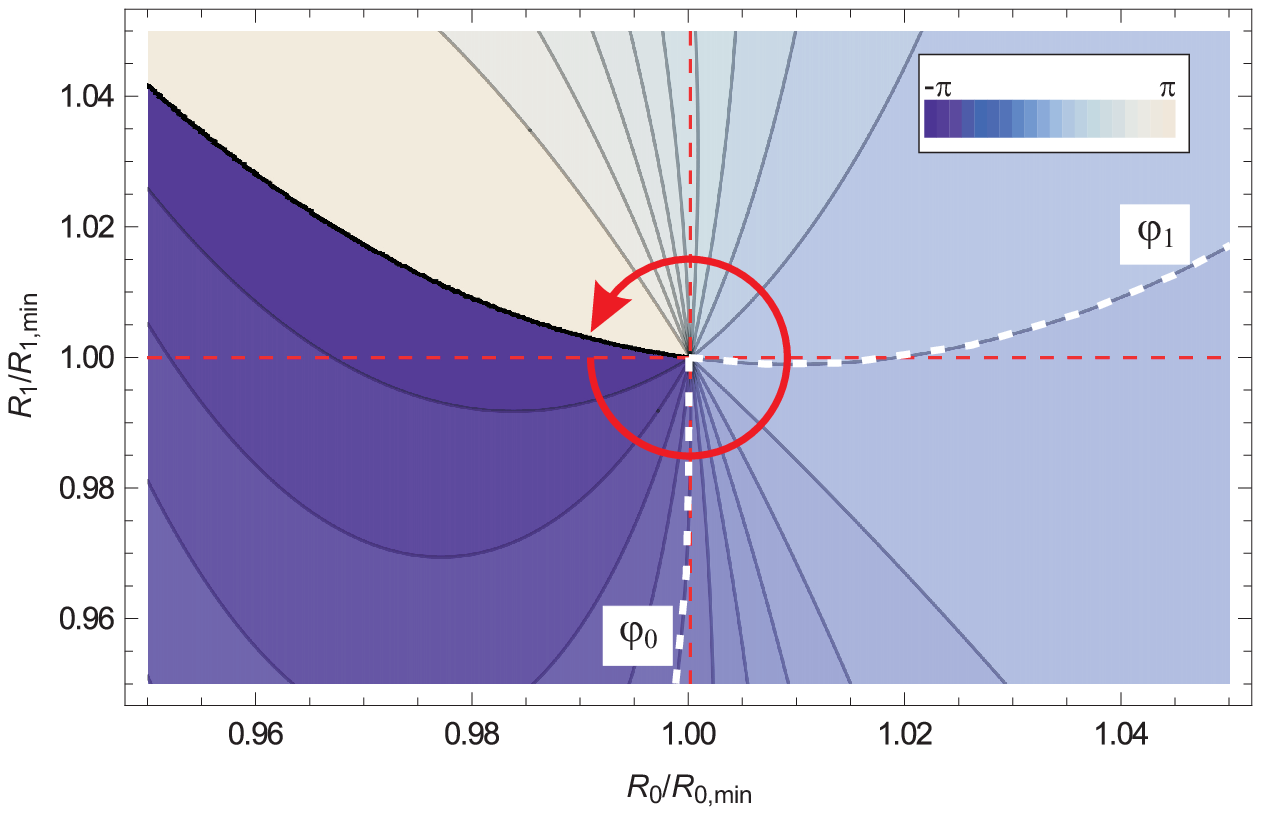}
	\label{fig:07-b}
	}
	\caption{
Contour plots of absolute value \subref{fig:07-a} and argument \subref{fig:07-b} of the effective spring constant $K_{\rm eff}$ of piezoelectric actuator shunted by the negative capacitor shown in Fig.~\ref{fig:02-NCscheme} as functions of the resistances $R_0$ and $R_1$. The values of $R_0$ and $R_1$ are normalized by values $R_{0,min}$ and $R_{1,min}$, respectively, which yield the zero absolute value of the effective spring constant $K_{\rm eff}=K^\prime_{\rm eff}+iK^{\prime\prime}_{\rm eff}$.  
}
\label{fig:07}
\end{figure*}

The reason for broadening the frequency range can be seen in Fig.~\ref{fig:05-broad}. Figures~\ref{fig:05-broad}(b) and (c) show the comparison of the measured frequency dependence of the electric impedance absolute value and phase of the piezoelectric actuator with the calculated values of the electric impedance of the negative capacitor with the narrow and broad frequency range reference capacitances  $Z_{1}$  shown in Fig.~\ref{fig:04-Z1AandB}(a) and (b). In Figs.~\ref{fig:05-broad}(b) and (c), it is seen that the electric impedances of the piezoelectric actuator and the negative capacitor reference capacitance are close to each other in the broad frequency range. Figure~\ref{fig:05-broad}(d) shows the frequency dependence of the real and imaginary parts of the effective Young\rq{}s modulus. It should be noted that the decrease in the Young\rq{}s modulus in broad frequency range results in a decrease in the resonant frequency of the system by approximately 400\,Hz. This yields an increase in the transmissibility of vibrations in sub-resonant frequencies.

\section{Adaptive system for the vibration isolation}
\label{sec:adaptive} 

An important issue accompanied with the vibration isolation system with manually adjusted negative capacitor is shown in Fig.~\ref{fig:06-TrTimeManVsAuto}. The solid line shows the time dependence of the vibration isolation efficiency in changed operational conditions in the system with manually adjusted negative capacitor. The vibration isolation system was turned on at a time ``1 minute\rq\rq{} and the decrease in the transmissibility of vibrations by the 20~dB was achieved. 
%
Then, the piezoelectric actuator was exposed to a slight heat irradiation from 100\,W bulb (tungsten lamp), which was placed in a distance of 25\,cm.
There is seen approximately 15 dB decrease in the suppression level of the transmissibility of vibrations [see Eq. (\ref{eq:13})] after 5 minutes. 

To avoid the severe deteriorative effect of changing operational conditions on the vibration isolation efficiency, the adaptive vibration isolation system has been implemented. The next Subsection describes the principle of the control algorithm.

\subsection{Iterative control law}
\label{subsec:adaptive_law} 

A simple control algorithm can be formulated using the analysis of contour plots of the absolute value and argument of, in general complex, effective spring constant $K_{\rm eff}$ of the piezoelectric actuator that is shunted by the negative capacitor. The negative capacitor is shown in Fig.~\ref{fig:02-NCscheme} as functions of resistances $R_0$ and $R_1$. Such plots with values $R_0$ and $R_1$ normalized by values $R_{0,min}$ and $R_{1,min}$, respectively, are shown in Fig.~\ref{fig:07}. The values $R_{0,min}$ and $R_{1,min}$ represent the optimal values of resistances in the negative capacitor that yield the zero absolute value of the effective spring constant. One can see that the absolute value of $K_{\rm eff}$ reaches zero for $R_0/R_{0,min}=1$ and $R_1/R_{1,min}=1$. A more interesting graph is shown in Fig.~\ref{fig:07}\subref{fig:07-b} for the argument of the effective spring constant $K_{\rm eff}=K^\prime_{\rm eff}+iK^{\prime\prime}_{\rm eff}$. One can see that the value of $\arg(K_{\rm eff})$ monotonically increases as the point $\left(R_0,\,R_1\right)$ goes around the optimal adjustment $\left(R_{0,min},\,R_{1,min}\right)$ in the counter-clockwise direction, as indicated by the arrow. 

Thus, one can immediately determine in which ``direction" is the optimal adjustment $\left(R_{0,min},\,R_{1,min}\right)$ with respect to its immediate value $\left(R_0,\,R_1\right)$ by measuring the argument of $K_{\rm eff}$, i.e. $\varphi=\arg(K_{\rm eff})$. Using this principle, it is possible to formulate the iterative control algorithm as follows:
\begin{subequations}
\label{eq:17_IterControlLaw}
\begin{eqnarray}
	R_{0,n+1}&=&
		\left\{
			\begin{array}{ll}
				R_{0,n}+\Delta R_0&\mbox{ for }\varphi<\varphi_0,\\
				R_{0,n}-\Delta R_0&\mbox{ for }\varphi_0<\varphi<\varphi_0+\pi,\quad\\
				R_{0,n}+\Delta R_0&\mbox{ for }\varphi_0+\pi<\varphi,
			\end{array}
		\right.\\
	R_{1,n+1}&=&\left\{
			\begin{array}{ll}
				R_{1,n}+\Delta R_1&\mbox{ for }\varphi<\varphi_1,\\
				R_{1,n}-\Delta R_1&\mbox{ for }\varphi>\varphi_1.
			\end{array}
		\right.
\end{eqnarray}
\end{subequations}
Symbols $R_{0,n+1}, R_{0,n}$ and $R_{1,n+1}, R_{1,n}$ are the ``new'' and ``old'' values of resistances $R_{0}$ and $R_{0}$, respectively. Values $\Delta R_0$ and $\Delta R_1$ are the resistance increments achievable in the negative capacitor. Symbols $\varphi_0$ and $\varphi_1$ stand for the critical values of $\arg(K_{\rm eff})$ indicated in Fig.~\ref{fig:07}\subref{fig:07-b}. The particular values of $\varphi_0$ and $\varphi_1$ should be usually determined experimentally. 

\begin{figure*}[t]
\begin{center}
    \includegraphics[width=0.8\textwidth]{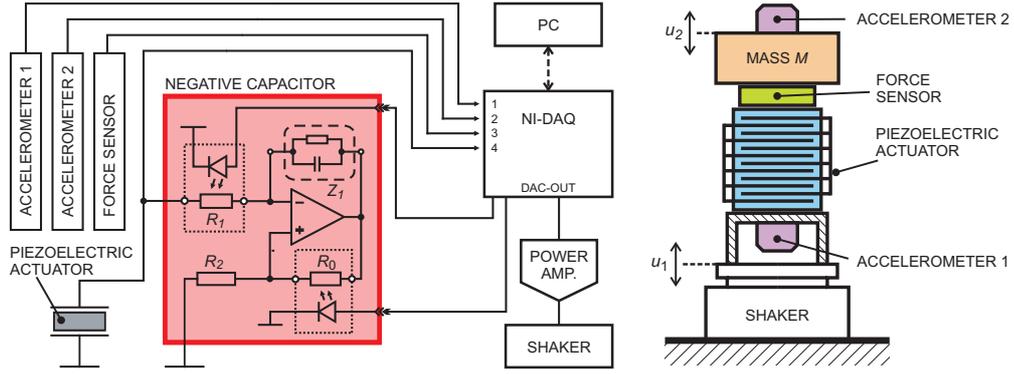}
\caption{
Combined system for the measurement of transmissibility of vibration through the adaptive vibration isolation system is shown on the right-hand side. The measurement part of the system consists of two accelerometers. The vibration isolation part of the system consists of a piezoelectric actuator shunted by an electronically adjustable negative capacitor. Electronic scheme of the combined system is shown on the left-hand side. Signals from accelerometers are used to calculate the transmissibility of vibrations. A signal from the force sensor and applied voltage from the negative capacitor are used for the estimation of the effective spring constant $K_{\rm eff}$ of the shunted piezoelectric actuator. The estimated value of the argument of $K_{\rm eff}$ is used for calculation of corrections to the values of resistances of electronically adjustable resistors $R_0$ and $R_1$.}
\label{fig:06-SchemeAdaptive}
\end{center}
\end{figure*}
%

In the next Subsection, the simple way for the estimation of the complex value of effective spring constant is presented.

\subsection{Estimation of the effective spring constant}
\label{subsec:adaptive_EstimKEff} 

The effective value of the spring constant is given by the ratio of the transmitted force $F$ through the piezoelectric actuator over it elongation $\Delta l$, see Eq.~(\ref{eq:07}). The transmitted force can be easily measured using a piezoelectric force sensor. The actuator elongation can be estimated using the following idea. 

When the negative capacitor is close to its required optimal adjustment, the transmitted force through the piezoelectric actuator is very small. When the transmitted force $F$ is small, it follows from Eq.~(\ref{eq:05})  that first term on the right-hand-side of Eq.~(\ref{eq:05}), i.e. $\left(1/K_S\right)F$, is much smaller than the second term, i.e. $dV$. In this situation, the elongation of the piezoelectric actuator is dominated by the inverse piezoelectric effect and, thus, it is proportional to the voltage applied from the negative capacitor, i.e. $\Delta l\propto V$.

In order to estimate the argument of the effective spring constant, it is sufficient to calculate the phase difference between the signal from the force sensor $F$ and the voltage $V$ applied from the negative capacitor:
\begin{equation}
	\arg\left(K_{\rm eff}\right) \approx \arg F - \arg V.
	\label{eq:18}
\end{equation}

\subsection{Implementation of the adaptive vibration isolation system}
\label{subsec:adaptive_eaNC} 

The above described control algorithm has been implemented in the adaptive vibration isolation system, which is shown in Fig.~\ref{fig:06-SchemeAdaptive}. Due to implementation convenience, the data-acquisition part of the adaptive vibration isolation system was combined with the system for the measurement of vibration transmissibility. Nevertheless, these two systems were independent.

The adaptive vibration isolation system consists of a force sensor and a piezoelectric actuator shunted by an electronically controlled negative capacitor. The force sensor was realized as a piezoelectric plate with a charge amplifier Kistler 5015A. Such an arrangement requires a calibration, which is done prior experiments in the setup without the damping element. The transfer function of the force sensor is determined using a mass of the object and the signal from the output accelerometer. This is simple and fast arrangement that allows precise force measurements up to high frequencies. The signal from the force sensor and applied voltage from the negative capacitor are used for the estimation of the effective spring constant $K_{\rm eff}$ of the shunted piezoelectric actuator. The estimated value of the argument of $K_{\rm eff}$ is used for the calculation of corrections to values of resistances of electronically adjustable resistors $R_0$ and $R_1$ according to Eqs.~(\ref{eq:17_IterControlLaw}).

\begin{figure}[t]
\begin{center}
    \includegraphics[width=82mm]{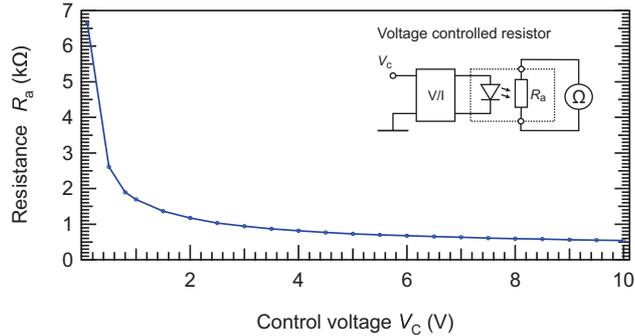}
\caption{
Example of the measured volt-ohm characteristics of the electronically adjustable resistor, which is constructed as a pair of a light-emitting diode and a photoresistor. The voltage $V_C$ controls the current through the diode and, therefore, the intensity of the emitted light, using a voltage-to-current converter. The intensity of the generated light controls the resistance $R_a$ of the photoresistor.}
\label{fig:07-EContrResistChar}
\end{center}
\end{figure}

In order to make the electronic control of resistances $R_0$ and $R_1$ in the negative capacitor possible, the manually adjusted trimmers were replaced by electronically controlled resistors, which were implemented as a pair of a light-emitting diode and a photoresistor. Example of the measured volt-ohm characteristics of the electronically adjustable resistor is shown in Fig.~\ref{fig:07-EContrResistChar}. The voltage $V_C$ controls the current through the diode and, therefore, the intensity of the emitted light, using a voltage-to-current converter. The intensity of the generated light controls the resistance $R_a$ of the photoresistor.

Instantaneous values of the incident and transmitted vibrations are measured by piezoelectric accelerometers PCB-352. These accelerometers have the resonant frequency at 40 kHz, which ensures a flat and phase correct transmission function in the frequency range of our experiments. Signals from the accelerometers are amplified by ICP amplifier. Electric signals from the accelerometers 1 and 2, force sensor, and the electric voltage applied to the piezoelectric actuator from the negative capacitor are measured and digitized by the data acquisition card NI PCI-6221. It should be stressed that accelerometers are a part of the measurement system only. They are used to measure the transmissibility of vibrations and to evaluate the efficiency of the adaptive vibration isolation system. Accelerometers are not used for the control of the vibration transmission. Signals from accelerometers do not enter the negative capacitor shunt and they are not used in the iterative control law.

Personal computer (PC) is used for three independent (but simultaneous) operations: First, it is used for the generation of the signal of the incident vibrations. In the Matlab software, a pseudo-random signal with few dominant harmonic components is generated. The output signal from the PC is introduced to the high-voltage amplifier and fed to the piezoelectric shaker. Second, PC processes signals from accelerometers and calculates the frequency dependence of the transmissibility of vibrations. Third, PC processes the signals from the force sensor and from the negative capacitor and generates control signals for electronically adjustable resistors in the negative capacitor according to the iterative control law.

The transmissibility of vibrations with harmonic time dependence of frequency 2\,kHz through the adaptive vibration isolation system is shown using dashed line in Fig.~\ref{fig:06-TrTimeManVsAuto}. It is seen that the transmissibility of vibrations remained constant even under varying operational conditions (i.e. ambient temperature). However, it should be noted that in the situation, where vibrations have harmonic time dependence, the estimation of effective spring constant argument is a straightforward and easy task. On the other hand, the consideration of harmonic vibrations greatly limits the applicability of the vibration isolation device. In order to eliminate this drawback and to broaden the applicability of the above described adaptive vibration isolation system, a modification of the control algorithm is necessary. This is described in the next Subsection. 

\subsection{Suppression of vibrations with a general time dependence}
\label{subsec:adaptive_fourier} 

In real situations, the incident vibrations usually consist of the sum of several randomly changing dominant harmonic components. These harmonic components appear in the system due to the eigen-frequencies of mechanical parts or due to vibration of revolving mechanical parts. In order to suppress the vibration transmission between the vibrating mechanical parts in real industrial applications, following modification to the control algorithm was implemented. 

First, signals from the force sensor and voltage applied from the negative capacitor are measured. If the amplitude of the signal from the force sensor exceeds some arbitrarily chosen threshold, the Fast Fourier Transformation is applied to the time dependencies of the measured signals in order to obtain their amplitude and phase frequency spectra. Then the distribution of the vibration power along the frequency axis is analyzed and the dominant harmonic component with the greatest amplitude is found. This dominant harmonic component is selected to be suppressed. At the selected frequency of the dominant harmonic component, the phase difference between the dominant harmonic components in the signals from the force sensor and from the negative capacitor output is calculated. The calculated value of the phase difference is used for the iterative corrections of the values of resistances $R_{0}$ and $R_{1}$ according to Eqs.~(\ref{eq:17_IterControlLaw}). After application of corrections to values of resistances $R_{0}$ and $R_{1}$, new time-dependences of signals from the force sensor and the negative capacitor output are measured and the above steps are periodically repeated until the dominant frequency is suppressed in the force sensor signal under a measurable level. 

\begin{figure}[t]
\begin{center}
    \includegraphics[width=85mm]{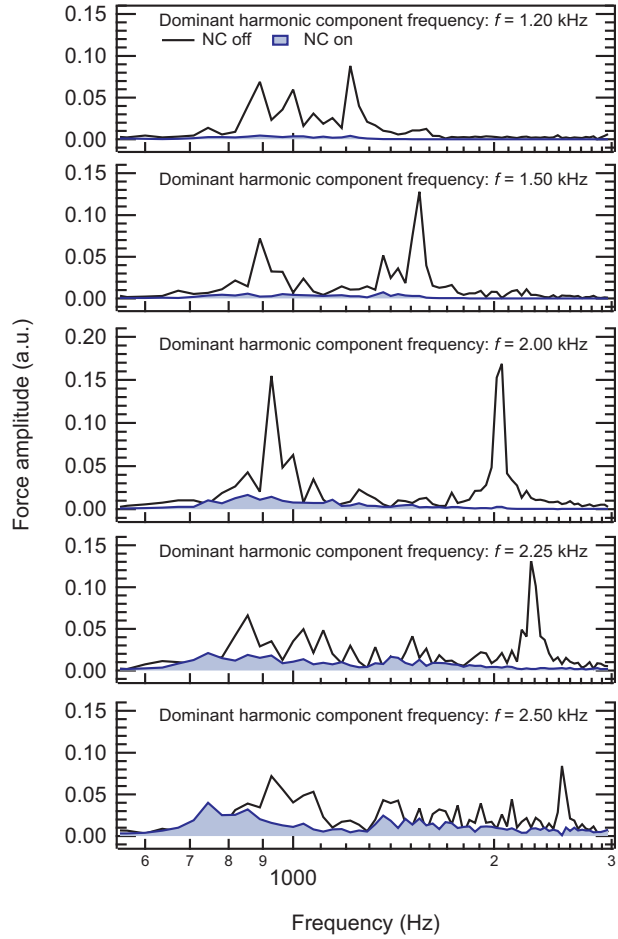}
\caption{
Spectra of the five force signals transmitted through the vibration isolation system with different frequencies of the dominant harmonic component. Solid black line indicates the force amplitude spectra transmitted through the piezoelectric actuator disconnected from the negative capacitor. The zero-filled solid blue lines indicates the amplitude spectra of the force transmitted through piezoelectric actuator shunted by the self-adjusted broad-frequency-range-matched negative capacitor. The vibration signal consists of a random noise and one dominant harmonic component of given frequency.}
\label{fig:08-Signals}
\end{center}
\end{figure}
%

\begin{figure}[t]
\begin{center}
    \includegraphics[width=85mm]{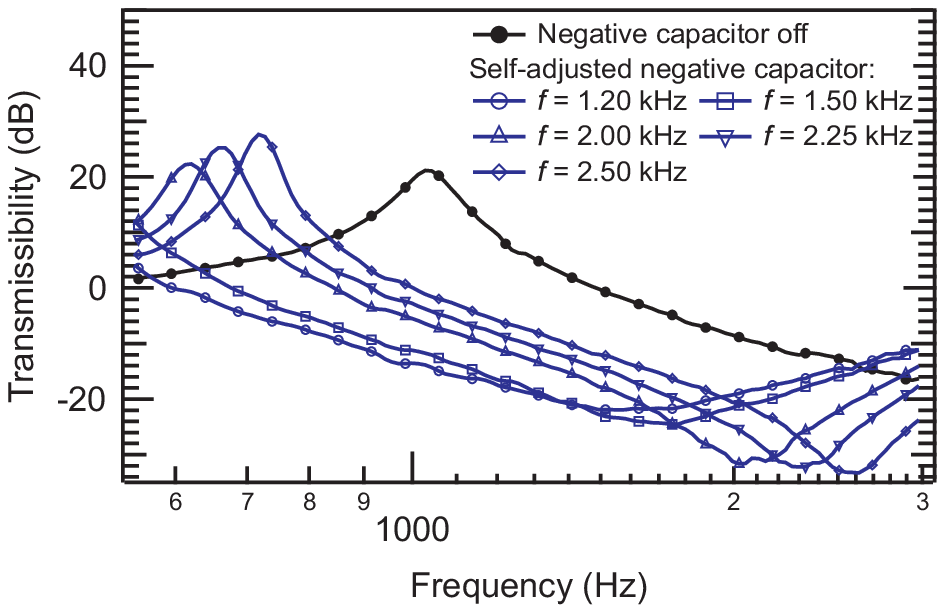}
\caption{
Frequency dependences of the transmissibility of vibrations through the piezoelectric actuator shunted by the adaptive broad-frequency-range-matched negative capacitor. Each curve corresponds the transfer functions of the adaptive system adjusted to cancel the vibration signal with the force amplitude spectra shown in Fig.~\ref{fig:08-Signals}. It should be noted that the decrease in the Young\rq{}s modulus in broad frequency range results in a decrease in the resonant frequency of the system by more than 300\,Hz. This yields an increase in the transmissibility of vibrations in sub-resonant frequencies.}
\label{fig:09-TrWideCompar}
\end{center}
\end{figure}
%

In order to evaluate the performance of the adaptive broad band vibration suppression device, five different vibration signals were generated and applied to the vibration isolation device. The vibration signal consists of a random noise and one dominant harmonic component of given frequency. Figure~\ref{fig:08-Signals} shows spectra of five force signals transmitted through the vibration isolation system with different frequencies of the dominant harmonic component. Solid black line indicates the force amplitude spectra transmitted through the piezoelectric actuator disconnected from the negative capacitor. The zero-filled solid blue lines indicate the amplitude spectra of the force transmitted through piezoelectric actuator shunted by the self-adjusted broad-frequency-range-matched negative capacitor.

The frequency dependences of the transmissibility of vibrations through the piezoelectric actuator shunted by the adaptive broad frequency range negative capacitor, which was self-adjusted to the five aforementioned vibration signals, are shown in Fig.~\ref{fig:09-TrWideCompar}. It is seen that the adaptive control algorithm adjusts the negative capacitor in such a way that the transmissibility of vibration curve has its minimum around the frequency of the dominant harmonic component in the vibration signal. Figure~\ref{fig:09-TrWideCompar} also indicates shifts of the mechanical resonant frequency of the system by more than 300\,Hz, which is due to the reduction of the effective spring constant of the piezoelectric actuator using the negative capacitor in the broad frequency range. This yields an increase in the transmissibility of vibrations in sub-resonant frequencies. Such an unwanted phenomenon can be easily eliminated by inserting high-pass filter in the negative capacitor.

Finally, it should be noted that the iterative control algorithm can conveniently compensate the effects of dielectric nonlinearity of piezoelectric actuator. The point is that with an increase in the amplitude of the incident vibrations, the voltage applied from the operational amplifier is also increased (more or less proportionally to the increase in the incident vibration amplitude). Then, the permittivity (and capacitance) of the piezoelectric actuator seen from the negative capacitor is slightly changed due to a dielectric nonlinearity (usually according to the Rayleigh law). This causes maladjustment of the system and unwanted drop in the efficiency of the vibration isolation. However, if the change in the amplitude of the incident vibration is not extremely fast, so that the system remains stable, the iterative control algorithm quickly compensates the changes in the piezoelectric actuator capacitance. The same behavior is expected even at full voltage range of the actuator, which can be achieved in systems with a standard high-voltage amplifier presented e.g. in the work by Fleming and Moheimani \cite{Fleming2004}.

\section{Conclusions}
\label{sec:conclusions}

The theoretical model of the vibration transmission through a piezoelectric actuator shunted by a negative capacitor is presented. The model has been verified using experiments performed on a narrow frequency (pure tone) vibration isolation. By a proper modification of the reference capacitor in the negative capacitor, it was successfully demonstrated that it is possible to achieve the vibration transmission suppression by 20 dB in the broad frequency range (from 1~kHz to 2~kHz). The iterative control law for automatic adjustment of the negative capacitor was disclosed using analyzing the absolute value and argument of the effective spring constant of the piezoelectric actuator shunted by a negative capacitor. A method for real-time estimation of the effective spring constant argument in the vibration isolation system was presented. However, the adaptive system in its basic arrangement is applicable only to the suppression of vibrations with harmonic time dependences. In order to eliminate this drawback, more evolved signal processing was implemented. It was shown that the iterative control algorithm is applicable also to vibrations with a general time dependence.

The advantages of the presented system for the suppression of vibration transmission stem from its simple electronic realization using an analog circuit with an operational amplifier, broad frequency range of the efficiently suppressed vibrations from 0.5~kHz to 3~kHz, and a simple control law that allows applying the automatic corrections to the negative capacitor, so that the system can work under varying operating conditions. In addition, the presented adaptive system is an example of a general concept of adaptive piezoelectric shunt damping, which can be easily modified and applied to a variety of different types of piezoelectric actuators and other electroacoustic transducers. All in all, the presented realization of the vibration isolation device offers a solution for many real noise and vibration problems.

\section*{Aknowledgments}

This work was supported by Czech Science Foundation Project No.: GACR 101/08/1279, co-financed from the student grant SGS 2012/7821 Interactive Mechatronics Systems Using the Cybernetics Principles, and the European Regional Development Fund and the Ministry of Education, Youth and Sports of the Czech Republic in the Project No. CZ.1.05/2.1.00/03.0079: Research Center for Special Optics and Optoelectronic Systems (TOPTEC). The authors acknowledge Julie Volfová for reading the manuscript.

\bibliographystyle{ieeetr}
\bibliography{ieee-04-utf8}

\begin{thebibliography}{10}

\bibitem{Hagood1991}
N.~W. Hagood and A.~von Flotow, ``Damping of structural vibrations with
  piezoelectric materials and passive electrical networks,'' {\em Journal of
  Sound and Vibration}, vol.~146, pp.~243--268, Apr. 1991.

\bibitem{Moheimani2006}
S.~O.~R. Moheimani and A.~J. Fleming, {\em Piezoelectric Transducers for
  Vibration Control and Damping}.
\newblock 2006.

\bibitem{Tsai1999}
M.~S. Tsai and K.~W. Wang, ``On the structural damping characteristics of
  active piezoelectric actuators with passive shunt,'' {\em Journal of Sound
  and Vibration}, vol.~221, no.~1, pp.~1--22, 1999.

\bibitem{Petit2004.proc}
L.~Petit, E.~Lefeuvre, C.~Richard, and D.~Guyomar, ``{A broadband semi passive
  piezoelectric technique for structural damping},'' in {\em {Smart Structures
  And Materials 2004: Damping And Isolation}} ({Wang, KW}, ed.), vol.~{5386} of
  {\em {Proceedings of the Society of Photo-optical Instrumentation Engineers
  (SPIE)}}, pp.~{414--425}, {2004}.

\bibitem{Morgan2002}
R.~A. Morgan and K.~W. Wang, ``Active-passive piezoelectric absorbers for
  systems under multiple non-stationary harmonic excitations,'' {\em Journal of
  Sound and Vibration}, vol.~255, pp.~685--700, Aug. 2002.

\bibitem{Morgan2002.jva-tasme.124.77}
R.~Morgan and K.~Wang, ``{An active-passive piezoelectric absorber for
  structural vibration control under harmonic excitations with time-varying
  frequency, part 1: Algorithm development and analysis},'' {\em {Journal of
  Vibration And Acoustics-Transactions of the ASME}}, vol.~{124}, pp.~{77--83},
  {JAN} {2002}.

\bibitem{RefLast1}
J.~Q. Ming~Yuan, Hongli~Ji and T.~Ma, ``Active control of sound transmission
  through a stiffened panel using a hybrid control strategy,'' {\em Journal of
  Intelligent Material Systems and Structures}, vol.~23, pp.~791--803, 2012.

\bibitem{Behrens2005.sms.12.18}
S.~Behrens, A.~Fleming, and S.~Moheimani, ``{A broadband controller for shunt
  piezoelectric damping of structural vibration},'' {\em {Smart Materials \&
  Structures}}, vol.~{12}, pp.~{18--28}, {FEB} {2003}.

\bibitem{Niederberger2004.sms.13.1025}
D.~Niederberger, A.~Fleming, S.~Moheimani, and M.~Morari, ``{Adaptive
  multi-mode resonant piezoelectric shunt damping},'' {\em {Smart Materials \&
  Structures}}, vol.~{13}, pp.~{1025--1035}, {OCT} {2004}.

\bibitem{Fleming2003.sms.12.36}
A.~Fleming and S.~Moheimani, ``{Adaptive piezoelectric shunt damping},'' {\em
  {Smart Materials \& Structures}}, vol.~{12}, pp.~{36--48}, {FEB} {2003}.

\bibitem{Badel2006.jasa.119.2815}
A.~Badel, G.~Sebald, D.~Guyomar, M.~Lallart, E.~Lefeuvre, C.~Richard, and
  J.~Qiu, ``{Piezoelectric vibration control by synchronized switching on
  adaptive voltage sources: Towards wideband semi-active damping},'' {\em
  {Journal of the Acoustical Society Of America}}, vol.~{119},
  pp.~{2815--2825}, {MAY} {2006}.

\bibitem{Date2000}
M.~Date, M.~Kutani, and S.~Sakai, ``Electrically controlled elasticity
  utilizing piezoelectric coupling,'' {\em Journal of Applied Physics},
  vol.~87, no.~2, pp.~863--868, 2000.
\newblock NIC.

\bibitem{Mokry2003jul}
P.~Mokrý, E.~Fukada, and K.~Yamamoto, ``Noise shielding system utilizing a
  thin piezoelectric membrane and elasticity control,'' {\em Journal of Applied
  Physics}, vol.~94, no.~1, pp.~789--796, 2003.

\bibitem{Mokry2003dec}
P.~Mokrý, P., E.~Fukada, and K.~Yamamoto, ``Sound absorbing system as an
  application of the active elasticity control technique,'' {\em Journal of
  Applied Physics}, vol.~94, no.~11, pp.~7356--7362, 2003.

\bibitem{Imoto2005}
K.~Imoto, M.~Nishiura, K.~Yamamoto, M.~Date, E.~Fukada, and Y.~Tajitsu,
  ``Elasticity control of piezoelectric lead zirconate titanate (pzt) materials
  using negative-capacitance circuits,'' {\em Japanese Journal of Applied
  Physics}, vol.~44, no.~9B, pp.~7019--7023, 2005.

\bibitem{Tahara2006}
K.~Tahara, H.~Ueda, J.~Takarada, K.~Imoto, K.~Yamamoto, M.~Date, E.~Fukada, and
  Y.~Tajitsu, ``Basic study of application for elasticity control of
  piezoelectric lead zirconate titanate materials using negative-capacitance
  circuits to sound shielding technology,'' {\em Japanese Journal of Applied
  Physics}, vol.~45, no.~9B, pp.~7422--7425, 2006.

\bibitem{Kodama2008}
H.~Kodama, M.~Date, K.~Yamamoto, and E.~Fukada, ``{A study of sound shielding
  control of curved piezoelectric sheets connected to negative capacitance
  circuits},'' {\em {Journal of Sound and Vibration}}, vol.~{311},
  pp.~{898--911}, {APR 8} {2008}.

\bibitem{Sluka2007}
T.~Sluka and P.~Mokrý, ``Feedback control of piezoelectric actuator elastic
  properties in a vibration isolation system,'' {\em Ferroelectrics}, vol.~351,
  pp.~51--61, 2007.
\newblock {8th European Conference on Applications of Polar Dielectrics
  (ECAPD-8), Metz, FRANCE, SEP 05-08, 2006}.

\bibitem{Preumont2008}
A.~Preumont, B.~de~Marneffe, A.~Deraemaeker, and F.~Bossens, ``The damping of a
  truss structure with a piezoelectric transducer,'' {\em Computers \&
  Structures}, vol.~86, pp.~227--239, {FEB} 2008.
\newblock {II ECCOMAS Thematic Conference on Smart Structures and Materials,
  Lisbon, PORTUGAL, JUL 18-21, 2005}.

\bibitem{Sluka2008}
T.~S. Sluka, H.~Kodama, E.~Fukada, and P.~Mokrý, ``Sound shielding by a
  piezoelectric membrane and a negative capacitor with feedback control,'' {\em
  IEEE Transactions on Ultrasonics, Ferroelectrics, and Frequency Control},
  vol.~55, pp.~1859--1866, {AUG} 2008.

\bibitem{Fleming2004}
A.~J. Fleming and S.~O.~R. Moheimani, ``Improved current and charge amplifiers
  for driving piezoelectric loads, and issues in signal processing design for
  synthesis of shunt damping circuits,'' {\em Journal of Intelligent Material
  Systems and Structures}, vol.~15, no.~2, pp.~77--92, 2004.

\end{thebibliography}

\end{document}